\documentclass[a4paper]{article}
\usepackage{authblk}
\usepackage{mathptmx}
\usepackage{xcolor}
\usepackage{amsmath}
\usepackage{geometry}
\usepackage{graphicx}\graphicspath{{images/}}
\usepackage[T1]{fontenc}
\usepackage{hyperref}
\usepackage[backend=biber, style=numeric, sorting=none]{biblatex}
\newcommand{\pa}{\partial}
\begin{document}
\large
\title{Coarse grained modeling of self assembled DNA 3D structure using pragmatic soft ellipsoid contact potential}
\author[1]{Abhirup Das}
\author[2]{Jayashree Saha\thanks{Correspondence: jsphy@caluniv.ac.in}}
\affil[1, 2]{Department of Physics, University of Calcutta, 92 Acharya Prafulla Chandra Road, Kolkata 700009}
\date{\today}

\maketitle
\begin{abstract}
    In this paper, we employ the soft ellipsoid contact type potential (ECP) in a generic coarse grained model of DNA and study the renaturation process. The interactions between the  bases are modeled using our ECP type potential, along with other interactions to capture bending, dihedral and solvent effects implicitly. Structure and helix formation is primarily driven by non-spherical geometry of the bases and excluded volume effects. Simulation results demonstrate that two isolated strands in a random initial configuration undergo spontaneous helix formation during annealing. The model captures the qualitative features of the coil-helix transition as well as the morphology of the helical structure. However, the present parameterization of our model does not yet capture the proper right chirality and detailed helical geometry associated with DNA. Thus, its capability for making any quantitative predictions remains limited. Statistical and physical properties of our model are discussed in detail.
    \vspace{1em}
\end{abstract}

\section{Introduction}
DNA is one of the most important molecules, required for the propagation of life. Despite its apparent complex structure, it is built from chain of repeating units consisting of three main moieties, a phosphate group, a deoxyribose sugar, and one of four types of bases. The four bases, namely adenine, thymine, guanine and cytosine can be grouped into purines and pyrimidines based on their molecular structure. In standard DNA most commonly found in living cells, a purine molecule bound to a sugar backbone strand, takes part in hydrogen bonding interaction with a complementary pyrimidine~\cite{Saenger1995}. Under normal conditions, this purine-pyrimidine binding is what gives double-stranded DNA stability and their complementary nature ensures that replication of DNA is possible. This ability of replication and transcription of genetic information has made DNA a pivotal molecule for the growth of life.~\cite{Ferry2019}. In recent times however, beyond its central biological role, DNA has also found applications in nanotechnology. The binding specificity of DNA which allows it to undergo sequence specific self-assembly, enables the design of nanoscale devices such as DNA tweezers~\cite{Seeman2017}. Hence, study of this elusive molecule warrants merit and has thus been the center of biological research for more than a century. Today, the chemical structure of DNA is more or less known, but to get a comprehensive picture of the structure property relationship important for biophysics and the related intermediate processes involved, requires us to resort to computer simulations. Additionally, to capture the structural and dynamical behavior of DNA in a reasonably expeditious way, computational models capable of balancing detail and efficacy are crucially needed~\cite{Mu2023}.

Atomistic models simulated with force fields such as AMBER and CHARMM provide high accuracy but are limited to short time-scales, while continuum models sacrifice finer molecular detail. Coarse-grained (CG) simulations offer a middle ground, and numerous models—ranging from early one-dimensional lattice models of Peyrard and Bishop to more sophisticated models such as MARTINI for DNA have been developed to study DNA mechanics and self-assembly~\cite{Ingolfsson2013}. The model of Peyrard and Bishop~\cite{Peyrard2004} is a simple one-dimensional model, with bases bound to its neighbors via harmonic springs. The bases were allowed one degree of freedom perpendicular to the springs, and interacted via a Morse potential with its minima at the zero of the coordinates. It was successful in replicating the dynamics of thermal denaturation of DNA and has been a topic of subsequent research. Later on, three-dimensional CG models of DNA were developed to gain better insights on dynamical aspects. Maciejczyk et.al.~\cite{Maciejczyk2000} created a three-site model by considering a virtual atom for each of the three moieties of DNA and constructed a potential of mean force to define inter-particle interactions for molecular simulation. The model was applied for the study of the transition between A and B forms of DNA.\@ The popular oxDNA model by Ouldridge et.al.~\cite{Ouldridge2011}, the 3SPN model by Knotts et.al.~\cite{Knotts2007} and its subsequent extensions~\cite{Sambriski2009}~\cite{Hinckley2013} and the MARTINI CG model extended to DNA~\cite{Uusitalo2015} are popular coarse grained models extensively used for biophysical studies of DNA, and hence are part of molecular simulation packages. These models use spherical geometry for the bases with directionally modulated potentials to account for asymmetry in base interactions. Other models such as the model by Morris-Andrews et.al.~\cite{MorrissAndrews2010} and Li et.al.~\cite{Li2016} used ellipsoidal bases interacting via anisotropic Gaussian overlap potential (GOP) like  potentials to better capture the planar geometry of the bases and their resulting interactions. Such potentials give an advantage over spherically symmetric potentials as the concept of orientation is woven into the mathematical form itself.

In this paper, we employ the soft ellipsoid contact type potential (ECP), to model the non-spherical geometry and interactions between bases in a generic coarse grained model of DNA. Previously, the soft ECP type potential has been applied in the study of the self-assembly of lipid bilayer phases~\cite{Bhowmick2025} with considerable success. The model potential is an extension of the ECP potential \cite{Perram1996}\cite{Paramonov2005}\cite{Ayton1995} with suitable modification to account for direction dependent energy scale correction~\cite{Saha2011}\cite{Saha2016}. In the next section, we present the coarse grained model, analyze the resulting structures in the subsequent section, and discuss some future prospects.
\section{Model}

The potential model is composed of six terms. There exist two terms denoted by \(V_{bend}\) and \(V_{dihedral}\) for the bonded interactions which account for the covalent structure of the molecule. The non-bonded interaction terms \(V_{ecp}\) and \(V_{solv}\) account for base pairing and implicit solvent interaction for renaturation of DNA. \(V_{exc}\) is the excluded volume interaction. Long range electrostatics is included via the term \(V_{el}\). The terms are explained in more detail below. 

\[
  V = V_{bend} + V_{dihedral} + V_{exc} + V_{el} + V_{ecp} + V_{solv}
\]

The covalent bonding interactions are usually modeled by a harmonic or FENE spring. In our work, however, we do not consider that degree of freedom. We instead use hard constraint implemented via the RATTLE algorithm to constrain the bonds to their equilibrium distances. The order of magnitude for bond vibrational modes in real biomolecular systems is higher than bending vibrational modes, so it is a valid assumption to take it as a constant with respect to the slower bending and dihedral modes of the system.

The bending interaction is modeled as a harmonic interaction
\begin{equation}
  V_{bend} = \sum_{i}^{angles} \frac{k_\theta}{2}(\theta_i - \theta_{0,i})^2
\end{equation}
Here, \( k_\theta \) is the force constant which signifies how flexible the respective bond angle is about its equilibrium value of \(\theta_0\). Torsional potential is included using a first order cosine form potential
\begin{equation}
  V_{dihedral} = \sum_{i}^{dihedrals} \frac{k_\phi}{2}[1-\cos(\phi_i- \phi_{0,i})]
\end{equation}
Here, \(k_\phi \) is the force constant for the dihedral degrees of freedom and \(\phi_0\) is the respective equilibrium angle. These represent the three and four body interactions which are critical for capturing the conformational flexibility of the DNA backbone. The bending potential is primarily applied to straighten the backbone chain and between bases to promote better stacking efficiency. Let \(N_{S}\) and \(N_B\) be the total number of sugar and base beads on both strands respectively. The bending potential is applied to the angle between adjacent sugar (S) atoms \(S_{i}\), \(S_{i+1}\) and \(S_{i+2}\) on each backbone chain for all \(i=\left(1,N_{S}-2\right) \). To increase stacking efficiency, it is also applied between adjacent base (B) atoms \(B_{i}\), \(B_{i+1}\) and \(B_{i+2}\) for all \(i=\left(1,N_{B}-2\right) \) separately for each strand. This biases the bases to align their flat sides on top of each other, reinforcing the alignment tendency of the anisotropic interaction. The dihedral potential is applied to the sugar base backbone. In order, the atoms involved in the interaction are \(B_{i}\), \(S_{i}\), \(S_{i+1}\) and \(B_{i+1}\) for all \(i=\left(1,N_{B}-1\right) \). Let \(v_{d1}=B_{i+1}-S_{i+1}\), \(v_{d2}=S_{i+1}-S_i\) and \(v_{d3}=S_i-B_i\) be the involved vectors for this interaction. The dihedral angle is therefore given by the formula
\begin{equation}
\cos\ \phi=\frac{\left(v_{d1}\times v_{d2}\right)\cdot\left(v_{d2}\times v_{d3}\right)}{\left|v_{d1}\times v_{d2}\right|\left|v_{d2}\times v_{d3}\right|}
\end{equation}
The equilibrium value for the bending interactions is set accordingly to promote the straight chain configuration. The dihedral equilibrium value is set to prevent the bases from freely rotating around the backbone axis. The value of the force constants are set heuristically in order to satisfy two competing requirements - maintaining sufficient structural stiffness to prevent collapse of the individual strands and maintain structural integrity of the duplex, while also allowing enough flexibility for proper conformational sampling during annealing and equilibration. Over-constraining the potentials would restrict the accessible phase space, kinetically trapping the system in local minima. The resulting helical geometry emerges from this balance in conjunction with excluded volume interactions. Also, we mention that since the dihedral potential is symmetric about the dihedral angle, there is no bias for specific handedness of the helix in our model. At present, formation of both left and right-handed helices is equally likely. The equations for the forces and torques for the bending and dihedral potential are available in standard textbooks on molecular simulation~\cite{Allen2017}~\cite{Rapaport2010}.

\(V_{exc}\) is the excluded volume potential added to all the virtual particles in the model. This interaction is primarily repulsive and applied between all the different atoms. However, base-base interactions and atoms with direct constraints between them are excluded from this interaction. Interactions involving two bases, whether same strand or different, are handled using a separate \(V_{ECP}\) potential, the details of which shall be discussed shortly. The excluded volume interaction is represented by a repulsive Weeks-Chandler-Andersen (WCA) type potential of the following form
\begin{equation}
    V_{exc} =
  \begin{cases}
    \displaystyle\sum_{i<j}^N\epsilon_e \left[ \left(\frac{\sigma_e}{r_{ij}}\right)^{12} - 2\left(\frac{\sigma_e}{r_{ij}}\right)^{6} \right] + \epsilon_e,&( r\leq \sigma_e ) \\
    0, & ( r>\sigma_e )
  \end{cases}
\end{equation}
where \(N=N_S + N_B\), \(\sigma_e\) and \(\epsilon_e\) are the relative excluded volume diameter and the interaction strength respectively, and \(r_{ij}\) is the inter-particle distance. For sugar atoms, we choose \(\sigma_e = 6.4\)~\AA~inspired from the equilibrium distance between the sugar atoms in the backbone~\cite{Ouldridge2011}. The excluded volume diameter for the bases interacting with sugars is a similar, purely repulsive interaction, but with an additional orientation dependence factored in due to the non-spherical shape of the base. Such cases of interactions between non-identical particles are handled by the soft ECP type potential. The value of \(\epsilon_e\) is set to \(1.0\epsilon_0\) uniformly for all cases.

Long range electrostatics is modeled via the Debye-H\"uckel term. 

\begin{equation}
    V_{el}=\sum_{i<j}^{N_{S}}\frac{q_i q_j}{4\pi\epsilon_0\epsilon_r r_{ij}}e^{-\left(\frac{r_{ij}}{\kappa_D}\right)}
\end{equation}

This term is an idealized approximation to model the non-ideal behavior of ionic solutions due to long range effects of electrostatic interactions. The potential is given in the form of a Coulumbic repulsion term modulated by an exponential decay, with a characteristic length scale \(\kappa_D\) known as the Debye length. This potential is applied to the \(N_S\) sugar beads in the model. For an electrolytic solution, Debye length is dependent on the ionic strength I of the solution via the following relation

\begin{equation}
    \kappa_D=\sqrt{\frac{\varepsilon_0\varepsilon_rk_B T}{2N_A e^2I}}
\end{equation}
Here \(\varepsilon_r\) is the dielectric constant of the solution, T is the absolute temperature, and \(e\) is the charge of an electron. For physiological conditions with 100 mM [Na\(^+\)] solution, the Debye length is approximately 1 nm~\cite{Hedley2024}. This corresponds to the high salt concentration regime ($>100$ mM), which is physiologically relevant. Consequently, our model is not applicable to low salt conditions.

The next term is \(V_{solv}\) which is used to model the cumulative effect of the solvent interactions of the DNA backbone in an implicit manner. This term was introduced by Sambriski et al.~\cite{Sambriski2009} in the 3SPN.1 force field. This term is added as an attractive interaction to the sugar atoms in the model.
\begin{equation}
  V_{solv} = \sum_{i}^{N_S^{(1)}}\sum_{j}^{N_S^{(2)}} \varepsilon_s \left[ 1 - e^{-\alpha (r_{ij} - r_0)} \right]^2 - \varepsilon_s
\end{equation}
The solvent interaction term uses a Morse potential with interactions characterized by an inter-particle separation term \( r_{ij} \) and energy well depth \( \varepsilon_s\). The equilibrium distance is represented by \( r_0 \). For each of the \(N_S^{(1)}\) sugar atoms in one strand, the potential acts between itself and all the other sugar atoms in the opposite strand \(N_S^{(2)}\). For our simulation, we set the equilibrium distance at 20~\AA~which is the experimentally observed helix diameter for the B-DNA conformer. The width \(\alpha=5.333\)~\AA\(^{-1}\) is same as in the literature for the 3SPN.1 force field \cite{Sambriski2009} with \(\varepsilon_s = 1.0\epsilon_0\) taken heuristically as a constant for a reference solvent since in this work we do not study salt effects. This interaction strength should depend on the nature of the solvent environment for a more faithful representation.

\begin{figure}[!hb]
    \centering
    \includegraphics[width=0.5\columnwidth]{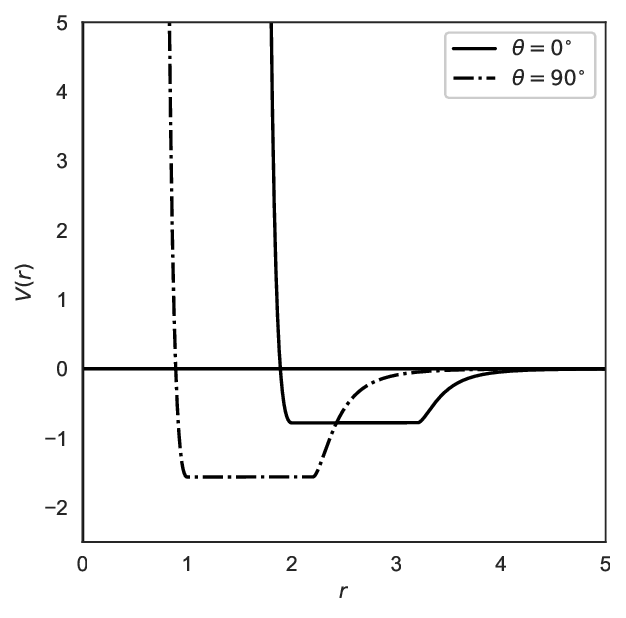}
    \caption{Potential energy profiles for the flat bottom implicit solvent potential for side-side \( \theta = 90^\circ\) and end-end configurations \(\theta=0^\circ \).}
    \label{fig:interaction}
\end{figure}

The last term \(V_{ecp}\) is the potential that handles the interactions between bases in our model. The bases in real DNA have planar ringed structures, whose interactions can be accurately captured by the soft ECP type potential. To represent the potential, we start with the shifted form of the Lennard-Jones (LJ) potential given by

\begin{equation}
  V_{ecp} = \sum_{i<j}^{N_B} \epsilon_{ecp} \left[ \left(\frac{\sigma_0}{r_{ij}-\sigma_{ecp}+\sigma_0}\right)^{12} - 2\left(\frac{\sigma_0}{r_{ij}-\sigma_{ecp}+\sigma_0}\right)^{6} \right]
\end{equation}

Here \(\sigma_0\) controls the width of the LJ potential, \(r_{ij}\) is the inter-particle distance, \(\sigma_{ecp}\) is the distance of closest approach and \(\epsilon_{ecp}\) is the respective energy well depth. For an ellipsoidal particle, the latter two quantities depend on both the orientations of the respective interacting particles and their direction of approach. The functional form we use for these terms is given by
\begin{align}
  \epsilon_{ecp} &= \epsilon_0 \epsilon_1^\nu(\hat{\alpha}, \hat{\beta}) \epsilon_2^\mu( \hat{\alpha}, \hat{\beta}, \hat{r}_{ij} ) \\
  \sigma_{ecp} &= \sigma(\hat{\alpha}, \hat{\beta}, \hat{r}_{ij})
\end{align}
Here \(\hat{\alpha}\) and \(\hat{\beta}\) are the orthogonal rotation matrices that define the orientation of the respective interacting particles in the lab frame. The form of \(\epsilon_{ecp}\) is similar to the form used in Gay-Berne potential~\cite{Gay1981}. The parameter values \( \nu = 2\) and \(\mu = 1\) are set as in Luckhurst et al.~\cite{Luckhurst1990}. The detailed form of the potential used, along with the expressions for the forces and torques are given in Appendix A.

We also employ two additional modifications of the base potential form. First, we extend the range of the potential with a flat bottom part. The parameter that controls the width of this flat bottom region is given by \(\omega_f\). Let~\( \rho=(r_{ij}-\sigma_{ecp}+\sigma_0)/\sigma_0\) be the dimensionless range parameter in the potential. Then
\begin{equation}
    \rho = \begin{cases}
        (r_{ij}-\sigma_{ecp}+\sigma_0)/\sigma_0 & (r_{ij} < \sigma_{ecp}) \\
        1 & (r_{ij} \geq \sigma_{ecp} \cap r_{ij} \le \sigma_{ecp} + \omega_f)\\
        (r_{ij}-\sigma_{ecp}+\sigma_0-\omega_f)/\sigma_0 & (r_{ij} > \sigma_{ecp} + \omega_f)
    \end{cases}
\end{equation}

The form of the potential for axial and transverse orientation of the two interacting ellipsoids is given in Figure \ref{fig:interaction}. Together with \( V_{solv} \), it represents the implicit solvent in our model. Second, we separate the interaction potential into a pairing part and a stacking part. For each base on one strand, the pairing part handles its interaction with all bases on the opposite strand. The stacking part, on the other hand, handles the interactions with its immediate neighbor bases above and below the concerned base on the same strand. Both these interactions have a cutoff and width parameter \(\omega_f\) which we keep same for both parts at 50 \AA~and 1.2\(\,\sigma_0\) respectively. We employ a hard cutoff at the boundary without any shifting or smoothing protocol. Since the potential is short-ranged, the jump in the potential is negligible at the boundary. The cutoffs are chosen sufficiently large to ensure proper energy conservation, allowing the bases to find and interact with each other.

\begin{table}[!b]
    \centering
    \caption{Parameter values used in the model.}
    \begin{tabular}{|c|c|}
        \hline
        Parameter & Value \\
        \hline\hline
        \( \sigma_0 \) & 1.0 \AA \\
        \( k_\theta \) & \(10.0\epsilon_0\) \\
        \( k_\phi \) & \(1.0\epsilon_0\) \\
        \(\theta_0\) & \(180^\circ\) \\
        \(\phi_0\) & \(0^\circ\) \\
        \( \sigma_e \) (Sugar) & \(6.4\sigma_0\) \\
        \( \sigma_e \) (Base) & \((6.8, 6.8, 3.4)\sigma_0\) \\
        \( \epsilon_e \) (Base) & \((1.0, 1.0, 2.0)\epsilon_0\) \\
        \( \epsilon_k \) & \(1.0\epsilon_0\) \\
        \( \varepsilon_s \) & \(1.0\epsilon_0\) \\
        \( \alpha \) & \(5.333\sigma_0\) \\
        \( r_0 \) & \(20.0\sigma_0\) \\
        \( \omega_f \) & \(1.2\sigma_0\) \\
        Cutoff & \(50.0\sigma_0\) \\
        \hline
      \end{tabular}
      \label{tab:parameters}
\end{table}

The simulation is set up in Fortran 90 programming language. We initialize two strands of 20 bases each by placing them at lattice points. The bases are represented by uniaxial ellipsoids and all bases have identical geometries \((\sigma_x, \ \sigma_y, \ \sigma_z) = (2.0,\ 2.0,\ 1.0)\, \times\,3.4\)~\AA. This choice is inspired by the standard values for the helix rise per base pair and helix diameter. The equilibrium stacking distance for real B-DNA is 3.4~\AA~\cite{Ouldridge2011}, hence we use this value for the axial diameter of the ellipse. The perpendicular diameters are set to twice the axial diameter to match the helix diameter approximately to the experimental value of 20~\AA. For the energies, we choose a similar ratio of \((\epsilon_x,\ \epsilon_y,\ \epsilon_z) = (1.0,\ 1.0,\ 2.0)\,\times\,\epsilon_0\). This choice  is consistent with experimental free energy ratios for stacking and pairing  which is approximately 1.5 to 2~\cite{Privalov2020}~\cite{AbrahamPunnoose2023}. The bond length between sugar atoms in the backbone is taken as 6.4~\AA~as discussed in Ref~\cite{Ouldridge2011}. The sugar-base bond is chosen to be 6.8~\AA~long from the sugar center to the base center to respect the helix diameter. The parameter values are heuristically chosen to be consistent with known experimental values for B-DNA. The values of all chosen parameters are given in Table \ref{tab:parameters}.

Initially the two strands were kept separate from each other. We use the Velocity Verlet integrator to solve the translational motion. For rotational motion, quaternion algebra has been used. The rotational integrator is set up in a fashion similar to Velocity Verlet, following the scheme of Rozmanov et al.~\cite{Rozmanov2010}. The simulation was run in the NVT ensemble using Langevin stochastic thermostat. The strength of the damping term was set to \(1.0\) in units of inverse time. The thermostat used is coupled to both the translational and rotational degrees of freedom. To prevent the strands from drifting too far from each other, a harmonic flat bottom symmetric confinement potential is used with its zero at the center of the simulation box. The flat bottom potential has a radius of 50~\AA. The strands do not feel any restorative force as long as the center of mass of the strands lie within the flat bottom part of the potential. Beyond that, the harmonic force pushes the strands towards the center. As the temperature is gradually reduced and bases start to pair, the confinement potential is ramped down to zero and the duplex is allowed to evolve freely. The restoring force from this potential is added to each particle equally, since all particles have unit reduced mass. This ensures no additional torques enter the system. More discussion about this can be found in Appendix B. 

The system was initially heated to \(T^\ast = 0.9\) which took the system to a random configuration. After that, we annealed the system through 50 equidistant temperature steps up to \(T^\ast = 0.1\) and equilibrated the system for around \(10^6\) total steps for each temperature step. The timestep for the simulation was set at \(0.005\) which is small enough to prevent overlaps, yet large enough to allow the system to properly sample the available phase space. The first 50\% of the total time steps was used for equilibration, and the final 50\% of time steps was used for production runs. For the production runs, we recorded the configuration state of the system such that for every temperature step, we obtained \(200\) equally spaced configurations in the time interval. The values of the potential energy, along with other thermodynamics parameters of the system are also recorded during the production runs. Reported average values for the thermodynamic quantities were obtained from single well equilibrated production runs at each temperature step.  Values for the other geometric observables reported, were obtained from the same trajectory at the lowest temperature point, far from the transition region. We further performed the simulation for five different initial configurations of the system and obtained similar results for the thermodynamic quantities.

\begin{figure*}[!b]
    \centering
    \includegraphics[width=0.4\textwidth]{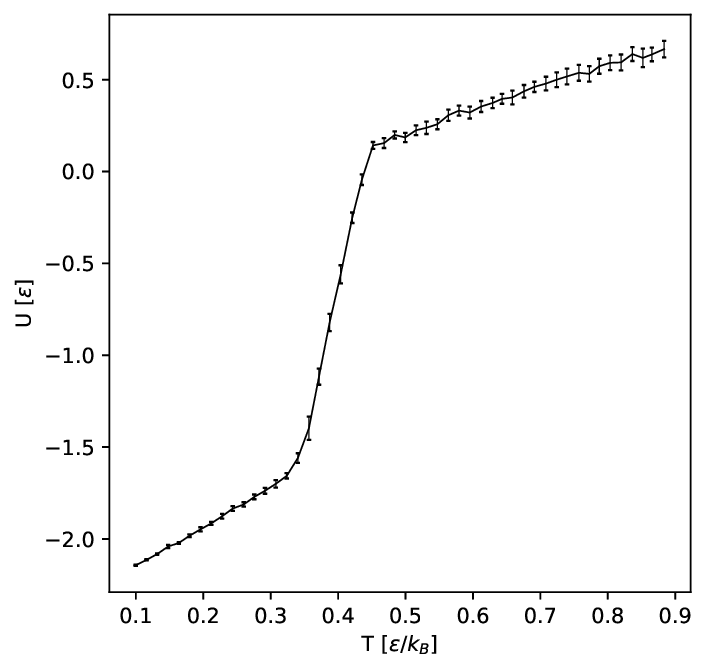}
    \includegraphics[width=0.51\textwidth]{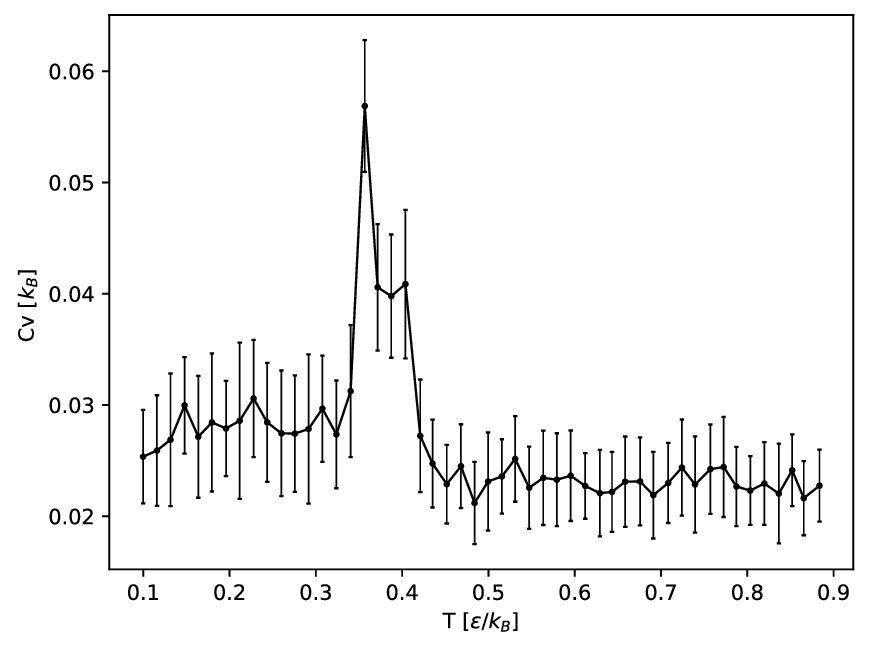}
    \caption{Phase transition curves obtained from simulation. (Left) Variation of the average CG potential (\(V_{total} / N\)) with temperature. There is a jump in the potential at around \(T^\ast = 0.35\) indicating a phase change where ssDNA hybridizes into dsDNA. (Right) Variation of specific heat capacity obtained from raw fluctuations in total energy \( \left[\langle E^2 \rangle - \langle E\rangle^2\right]/T^{\ast2} \) for each temperature. Fluctuations peak at around the same temperature where the drop occurs in Fig A, showing the transition behavior as the DNA strands alternate between hybridized and unhybridized states.}
    \label{fig:phasetransition}
\end{figure*}

\section{Results and Discussion}
\begin{figure}[!t]
    \centering
    \includegraphics[width=0.45\textwidth]{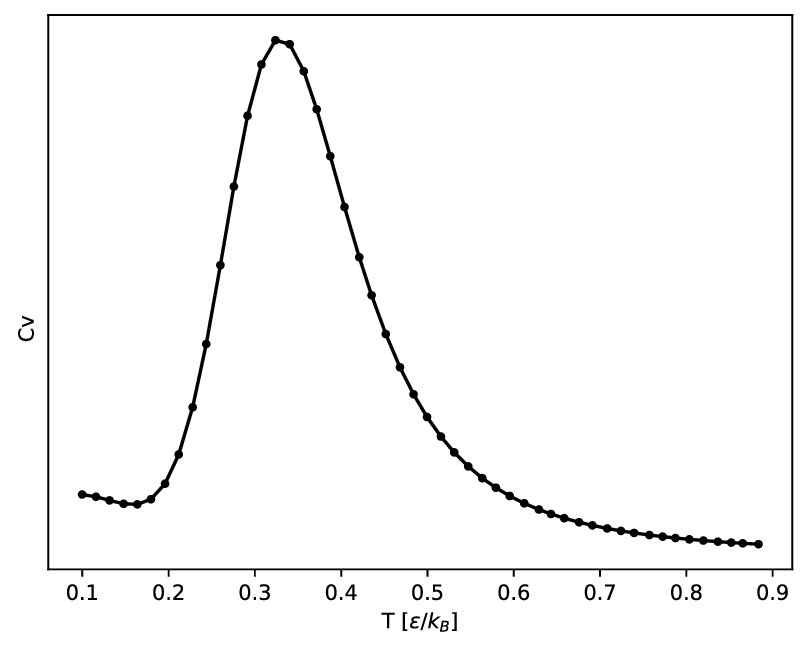}
    \caption{Specific heat graph obtained from MBAR method. The curve is in good agreement with the raw fluctuation data.}
    \label{fig:potmbar}
\end{figure}

Coarse-graining (CG) is a systematic dimensionality reduction technique that maps a high-dimensional atomistic system onto a simplified representation with fewer degrees of freedom. The two-body and higher-order CG interaction potentials effectively encode the many-body effects of the atomistic system. Our CG potential thus approximates the potential of mean force (PMF), with many-body effects integrated out according to the CG mapping scheme. A true PMF would require a rigorous parameterization scheme to ensure proper integration of degrees of freedom. However, for this work, we are mainly interested in capturing the hybridization behavior of the system. Thus, we use the total potential energy per particle  or the average CG potential energy (\(V_{total} / N\)) and track changes in its value with temperature.

In Figure \ref{fig:phasetransition}A, we plot the value of average CG potential energy with corresponding reduced system temperature. Since MD trajectories are essentially correlated in time, we
employ a block averaging scheme by dividing the total time in the production run into 50 blocks of equal size for each temperature point, ensuring that the width of each block is larger than the correlation time of the trajectory. The reported error bars in Figure \ref{fig:phasetransition} represent the standard error of the observable across the block means.  Close to the transition temperature, the correlation time increases, so we needed a longer trajectory and fewer blocks to get uncorrelated samples. We get a jump in the average potential energy at around \(T^\ast=0.35\) which indicates a large scale change in the system. After a while, the sharp drop plateaus into a gentle slope. We support the transition behavior from the trends in specific heat data. For specific heat, we try two methods. Firstly, we try a crude approach by plotting the raw fluctuation data obtained from the variance of the total energy function using the formula \( \left[\langle E^2 \rangle - \langle E\rangle^2\right]/T^{\ast2} \). This is shown in Figure \ref{fig:phasetransition}B. The plot shows a peak in the fluctuation data at around the same temperature. This is expected behavior since close to the transition point, system jumps between the single-stranded state and the double-stranded state. Due to the crude approach, our main peak gets a bit fragmented into a few bumps. However, the general trend is captured.

We also try a more statistically refined approach using the Multistate Bennet Acceptance Ratio (MBAR) method \cite{mbarchodera} to obtain the specific heat curve. The necessary requirement for MBAR method to work is that there is sufficient overlap between the energy distributions for each temperature point. Our temperature points are reasonably closely spaced, and we sufficiently equilibrate the system for every temperature point. Henceforth, we can safely assume there is considerable overlap between the energy distributions for MBAR method to work. We do this using the PyMBAR package in Python~\cite{mbargit}. The raw  energy data for each temperature point is fed to pymbar to obtain the specific heat curve. We take care to first decorrelate the data using the timeseries module in PyMBAR before passing it to the MBAR subroutine. This curve is shown in Figure \ref{fig:potmbar} which is in good agreement with our raw fluctuation data.

\begin{figure}[!t]
    \centering
    \includegraphics[width=0.45\textwidth]{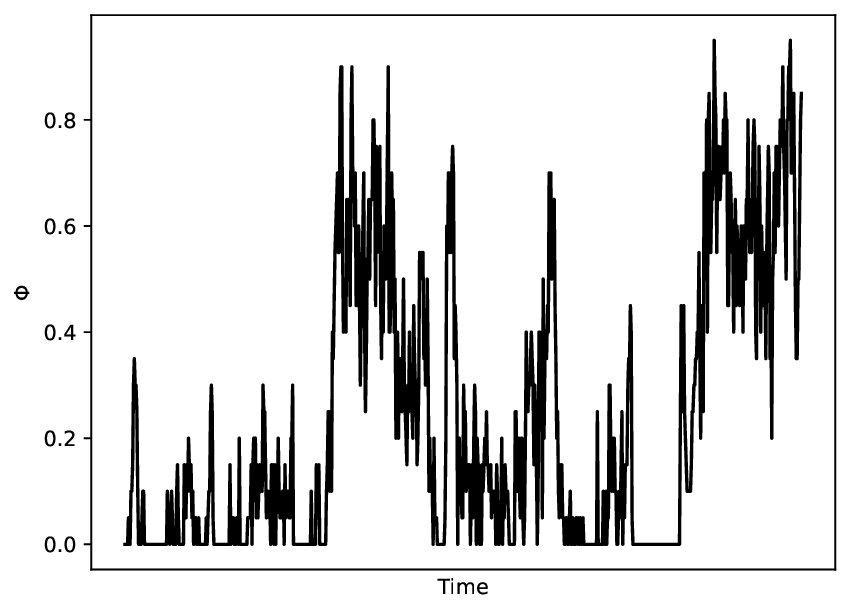}
    \includegraphics[width=0.45\textwidth]{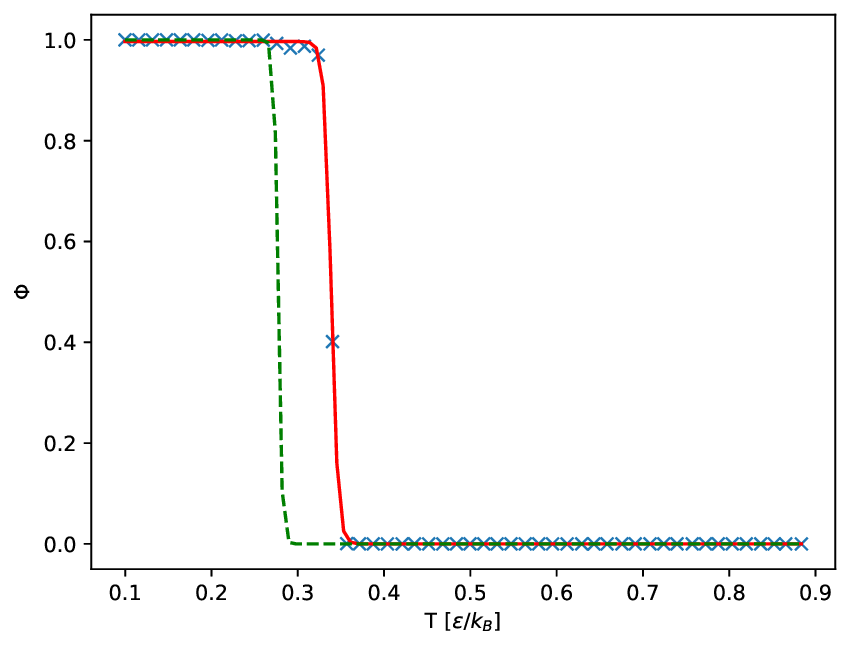}
    \caption{(Left) Variation of pairing fraction (\( \Phi \)) with time. The system jumps between paired and unpaired bases as the system cools towards hybridized state. Data is presented at around \(T^\ast=0.38\). (Right) Melting curve obtained by fitting the sigmoid function given in Equation \ref{eq:1} over the pairing fraction \( \Phi \) for each simulated temperature. The crosses \( (\times)\) are actual data from simulation over which the sigmoid function is fitted. The dotted curve denotes a reference experimental curve taken from Ref~\cite{Owczarzy2004} as mentioned in the text.}
    \label{fig:pairfrac}
\end{figure}

From a local perspective, as the temperature of the system decreases, the bases begin to pair. The strength of the cooperative stacking interaction affects, to an extent, the steepness of the transition. Figure~\ref{fig:pairfrac}A presents the fraction of paired bases
formed, also called the pairing fraction (\(\Phi\)) as a function of time for the system at \(T^\ast=0.38\). For this, we treat bases as paired if the distance between them is less than 12~\AA. Since the system is close to transition point, we get a lot of pairing and unpairing events between the bases. In  Figure~\ref{fig:pairfrac}B, we plot the pairing fraction \((\Phi)\) averaged over time, as a function of temperature. We try to fit a sigmoid curve over the pairing fraction data of the form
\begin{equation}
f\left(x\right)=\frac{a}{1+e^{-b\left(x-x_0\right)}}
\label{eq:1}
\end{equation}
The best fit curve is plotted in Figure \ref{fig:pairfrac}B. For reference, we also include an experimental melting curve from Ref~\cite{Owczarzy2004} for the 20-mer DNA sequence TACTTCCAGTGCTCAGCGTA at 69~mM [Na$^+$], which has a melting temperature of 333.45~K, shown as a dotted curve in Figure \ref{fig:pairfrac}B. We emphasize the fact that our results are meant to provide qualitative agreement with the experimental melting curve. We have not performed rigorous parameter fitting, since our model lacks a few key features of the real system. Exact matching of simulated and experimental melting temperatures for specific sequences and salt concentrations is beyond the scope of this work.

\begin{figure}[!t]
    \centering
    \includegraphics[width=0.45\columnwidth]{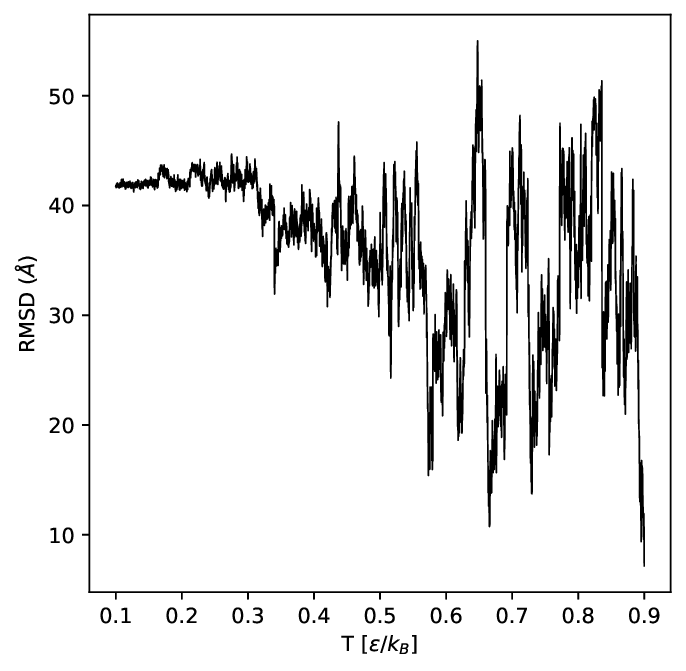}
    \includegraphics[width=0.45\columnwidth]{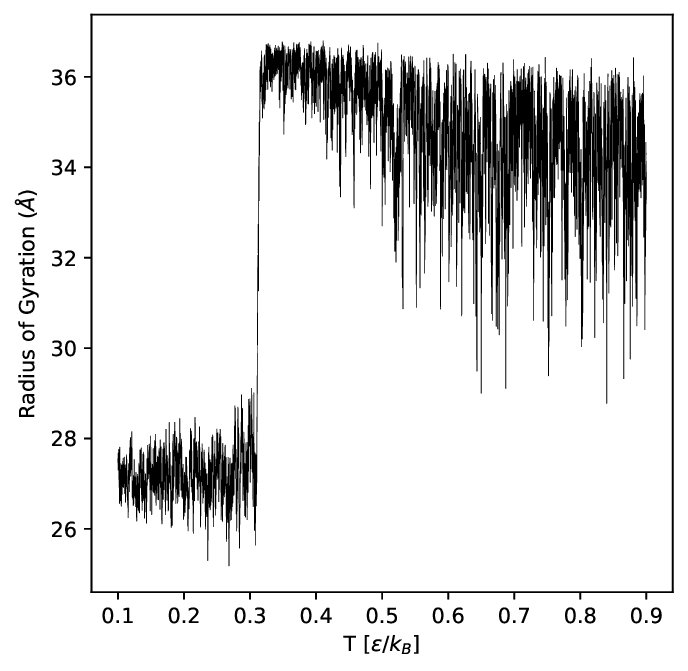}
    \caption{(Left) RMSD distribution showing the plateau behavior as the system hovers around the equilibrium ensemble. (Right) A plot of evolution of the radius of gyration of the system as a function of temperature. As RMSD begins to plateau, Rg jumps to a low value, as the extended chains coil into a single helix. Since we have a single helix, Rg is expected to be lower than the radius of gyration of the single stranded state. Since we store around \(200\) snapshots of the configuration across the production run for each of the \(50\) temperature points, we have a lot more data points than temperature points. Thus, the x scale used here is purely representative to portray the trend of the fluctuation across temperatures, rather than the actual value of the observable at that corresponding temperature.}
    \label{fig:rmsdrg}
\end{figure}

\begin{figure}[!b]
    \centering
    \includegraphics[width=0.45\textwidth]{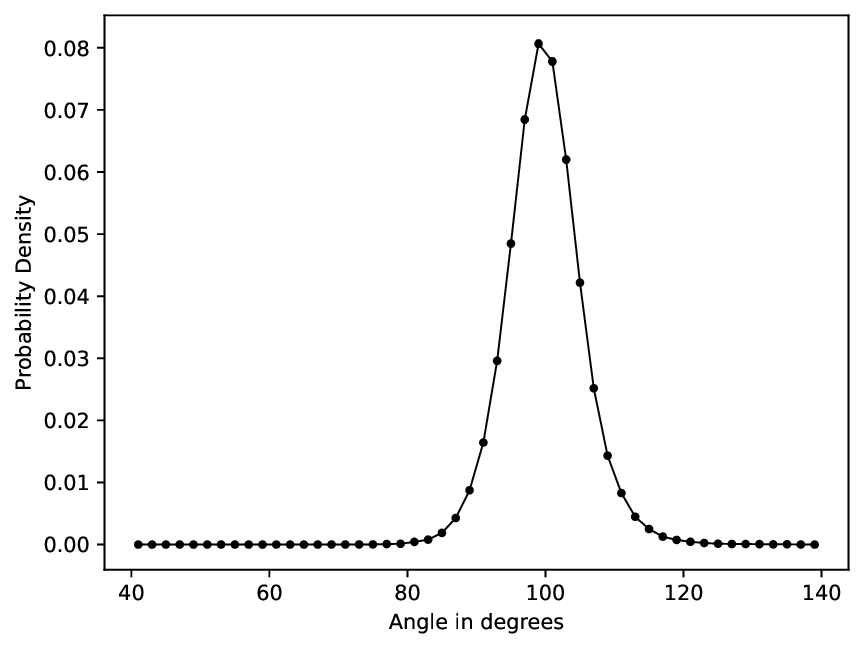}
    \includegraphics[width=0.45\textwidth]{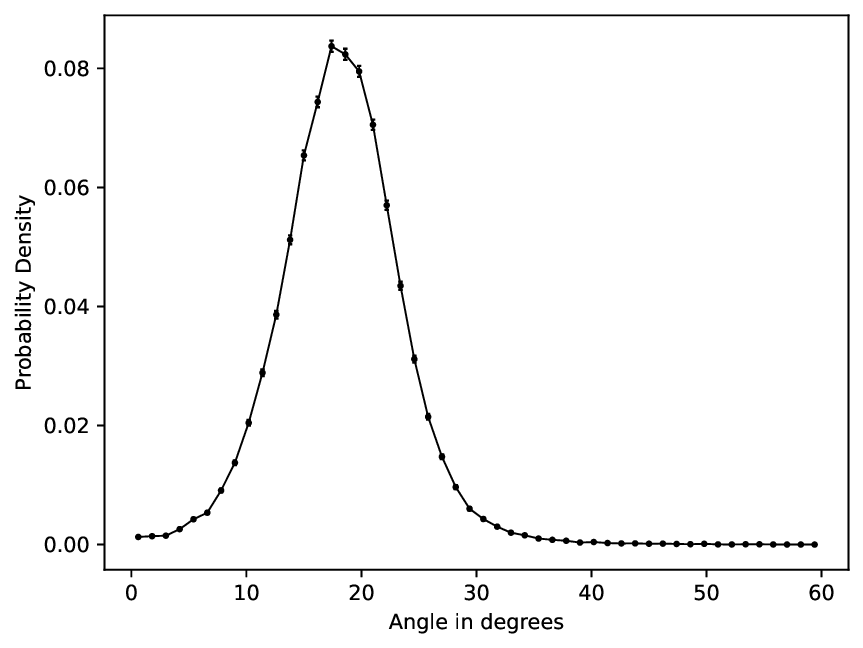}
    \caption{(Left) Bending angle distribution. (Right) Dihedral angle distribution. Both distributions settle at equilibrium value with peaks at around \(99^\circ\) and \(17^\circ\) respectively as the system cools towards hybridized state.}
    \label{fig:benddih}
\end{figure}

\begin{figure}[!t]
    \centering
    \includegraphics[width=0.45\textwidth]{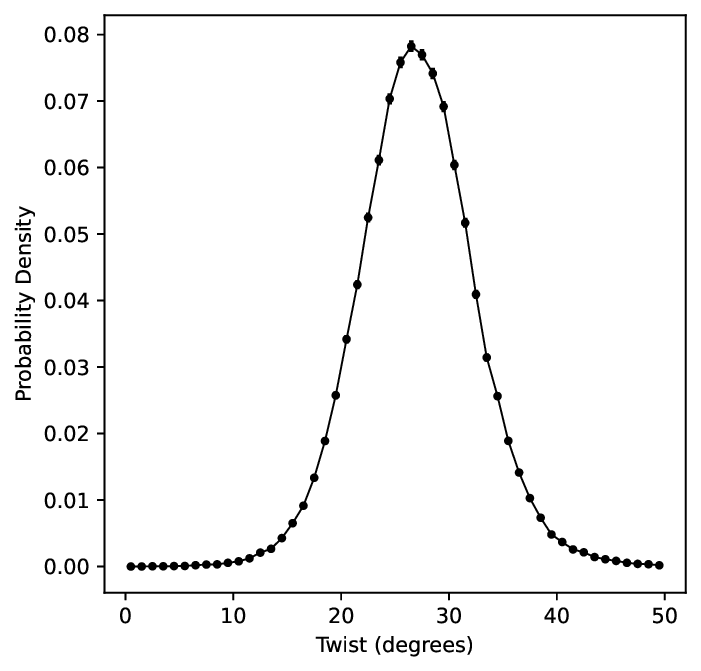}
    \includegraphics[width=0.45\textwidth]{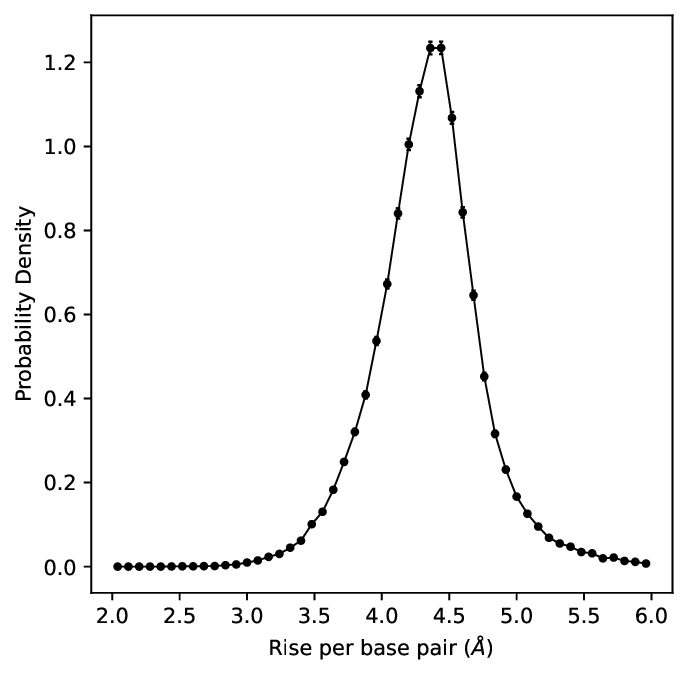}
    \caption{(Left) Twist angle obtained from simulation. (Right) Average rise per base pair obtained by constructing a local helical axis for each base according to the scheme as described in the text. The local helical axis is given by the principal component of the covariance matrix formed from the base centers around the bases being probed.}
    \label{fig:helixrise}
\end{figure}

In Figure~\ref{fig:rmsdrg}A, we plot the root mean squared deviation of the structures with time. The deviation is measured with reference to the initial randomized structure at time zero according to the formula
\begin{equation}
X_{\text{RMSD}}^t = \sqrt{\frac{1}{N}\sum_{i=1}^{N}\left(\text{X}^t_i - \text{X}^0_i\right)^2}
\end{equation}
We see that after an initial rise, the curve plateaus at a constant value. This indicates that our system has systematically drifted away from the initial reference structure and fluctuates around an equilibrium ensemble at lower temperatures. This reinforces the proper equilibration of our system and the validity of our annealing protocol. Figure~\ref{fig:rmsdrg}B shows the variation of the radius of gyration of the system. At the transition, the value of gyration radius drops sharply and settles at an equilibrium value lower than before, since the duplex state has a lower gyration radius than the single stranded state. The values obtained for individual structural parameters are reported in Table \ref{tab:values}.

The bending angle and dihedral angle distributions are shown in Figure~\ref{fig:benddih}. Figure~\ref{fig:benddih}A captures the distribution of the bending angle between the backbone and base, particularly the angle between \(S_{i}\), \(S_{i+1}\) and \(B_{i+1}\) for all \(i=\left(1,N_S-1\right) \). The equilibrium value reached is at around \(99^\circ\). The dihedral angle for the backbone bonds equilibrate at \(17^\circ\) in the hybridized state, as shown in Figure~\ref{fig:benddih}B. This value is equilibrated via stabilizing the pairing and stacking between the bases, respecting the excluded volumes of the surrounding atoms.

The observed value for twist angle obtained from simulations is around \(27^\circ\) which is plotted in Figure \ref{fig:helixrise}A. We also plot the distribution of average rise per base pair from our model in Figure \ref{fig:helixrise}B. To obtain it, we first construct a local helix axis around the bases being probed in the range of 3 bp. For example, let's say we want to find the rise between base index \(i\) and \(i+1\). So, we construct the local helix axis from base centers \(i - 3\) to \(i + 3\). The coordinates of the base centers are used to construct a local covariance matrix. The helix axis is obtained via principal component analysis (PCA), which is the eigenvector corresponding to the largest eigenvalue of the covariance matrix. The vector difference in positions \(d_1=B_{i+1}-B_i\) is projected onto the local helical axis to obtain the rise for bases \(B_i\) and \(B_{i+1}\). This distance averaged for all bases (excluding three bases from both ends) gives us the average rise per base pair. The estimated value from the graph is 4.37~\AA~which is far from the experimentally reported value of 3.4 \AA~\cite{Arnott1980}~\cite{Langridge1960}.

\begin{table}[!b]
    \centering
    \caption{Values of major structural quantities obtained from simulation.}
    \begin{tabular}{|c|c|c|}
        \hline
        Structural quantity & Value from simulation \\
        \hline\hline
        End to end distance (\AA) & \( 85.17 \pm 1.67 \) \\ 
        Radius of Gyration (\AA) & \( 27.11 \pm 0.44 \) \\ 
        Helix diameter (\AA) & \( 20.0 \pm 0.12 \) \\ 
        Rise per base pair (\AA) & \( 4.37 \pm 0.04 \) \\ 
        Bending angle (\(^\circ\)) & \( 99.68 \pm 0.75 \) \\ 
        Dihedral angle (\(^\circ\)) & \( 17.5 \pm 0.18 \) \\
        Twist angle (\(^\circ\)) & \( 27.1 \pm 0.12 \) \\
        Pitch Length (\AA) & \( 53.4 \pm 0.08 \) \\
        Persistence length (ssDNA) (nm) & \( 1.27 \pm 0.7 \) \\
        Persistence length (dsDNA) (nm) & \( 40.2 \pm 0.6 \) \\
        \hline
    \end{tabular}
    \label{tab:values}
\end{table}

As we can see, our model is unable to bring the observed twist angle and rise per base pair for B-DNA. Both the observed twist angle being smaller and the rise per base pair being larger than the experimental values (approx. 34$^\circ$ and 3.4 \AA~respectively) point to the same fact that our model has a larger helical pitch length (> 34~\AA), even though the helix diameter is correctly obtained. This is a limitation of our model. In our opinion, the discrepancy arises from the form of the implicit solvent Morse potential. This attractive potential is meant to facilitate structure formation to mimic the effects of surrounding solvent as in Ref~\cite{Sambriski2009}. However, it's designed in a way to keep one sugar strand a set distance away from the other. This hampers formation of unequal groove widths since the distance between alternate backbone strands itself is not a constant in the helix form. It alternates between larger (major grooves) and smaller (minor grooves) distances. Increasing the value of the twist to match experimental values in the presence of the implicit solvent potential leads to steric clashes, which is incompatible with the simplified 2-bead representation. Our model at present cannot account for unequal grooves, which is likely the reason why the obtained helical pitch is also larger than the experimentally observed value.

\section{Conclusion}

In this paper, we use the soft ECP type potential~\cite{Saha2011}\cite{Saha2016} in a coarse-grained model of DNA. The potential model is applied for base-base interactions. We separate the inter-strand base pairing interactions from intra-strand stacking interactions. This facilitates dsDNA formation by avoiding hairpin configurations. For the bonded interactions we stick to standard harmonic and cosine potentials for bending and dihedral interactions respectively. The bonds use hard constraints implemented via RATTLE algorithm. An excluded volume WCA potential is applied to prevent atoms from overlapping, the radius of exclusion sphere taken different for each type of particle.  We introduce implicit solvent effects via a Morse potential and a flat bottom part in the ECP type potential. Electrostatics is considered using Debye-H\"uckel term. This approximation is valid in the high salt concentration regime ($>100$ mM), which is physiologically relevant.

Proper handling of base geometry and accurate determination of excluded-volume effects as it interacts with the other beads present in the system is crucial for correctly obtaining the helical structure and overall morphology during DNA renaturation. For this purpose, we chose the soft ECP type potential for base-base and base-sugar interactions, the latter of which is purely repulsive. This is due to two useful properties of the potential that serves our requirements well. Firstly, it can accurately estimate the distance of closest approach for non-identical interacting particles for any given orientation and geometry. Secondly, it can correctly estimate the energy variation for cross interactions between two non-identical interacting particles. Instead of relying on predefined mixing rules, the framework uses a parametric optimization strategy to estimate the mixing parameters from the individual energy wells and orientational geometry of the interacting particles. Due to this feature, we only need to decide the excluded diameters and energy well depths for the identical bases. This makes the model potential more general and applicable to a wide range of systems, from inclusion of sequence effects to the study of DNA-protein and DNA-lipid complexes. The challenge that remains is to parameterize the potential for each of the four types of bases and the related interactions. We shall address this in our future  work.

\begin{figure*}[!b]
    \centering
        \fbox{ \includegraphics[width=0.4\textwidth]{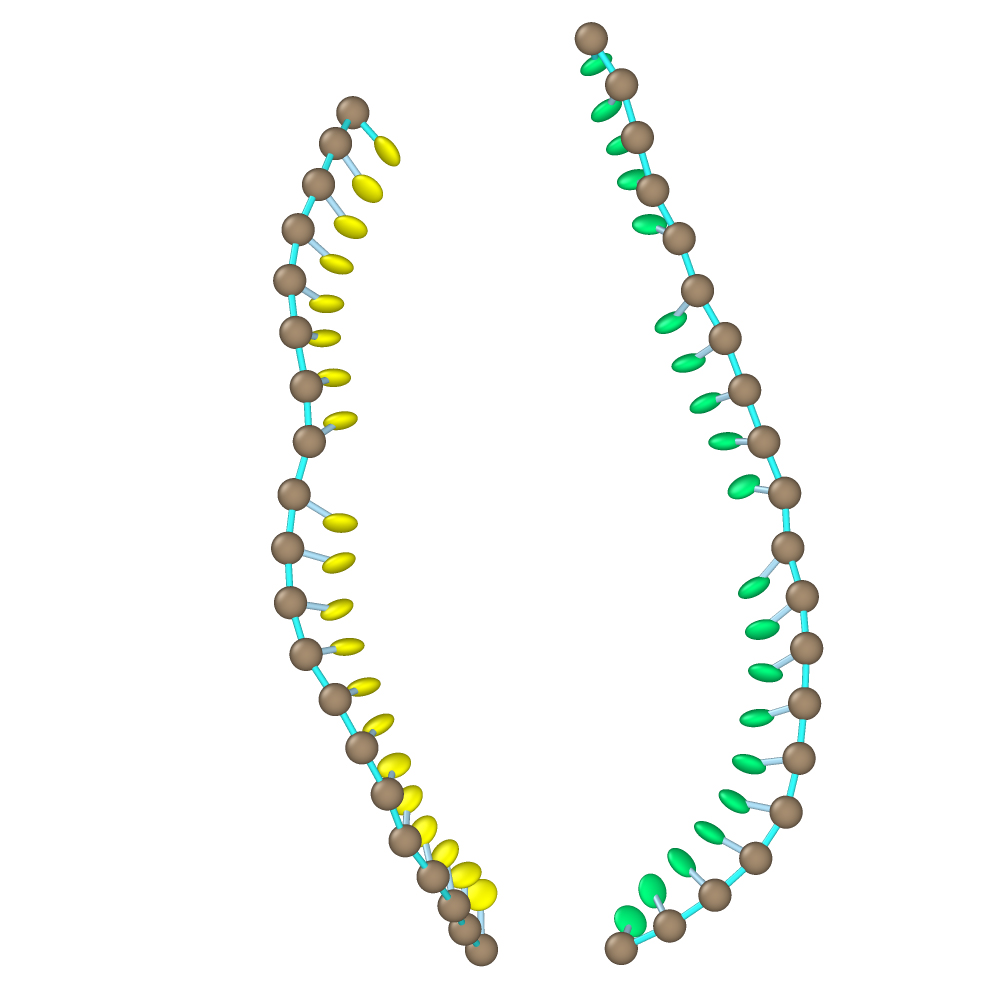}}
        \fbox{ \includegraphics[width=0.4\textwidth]{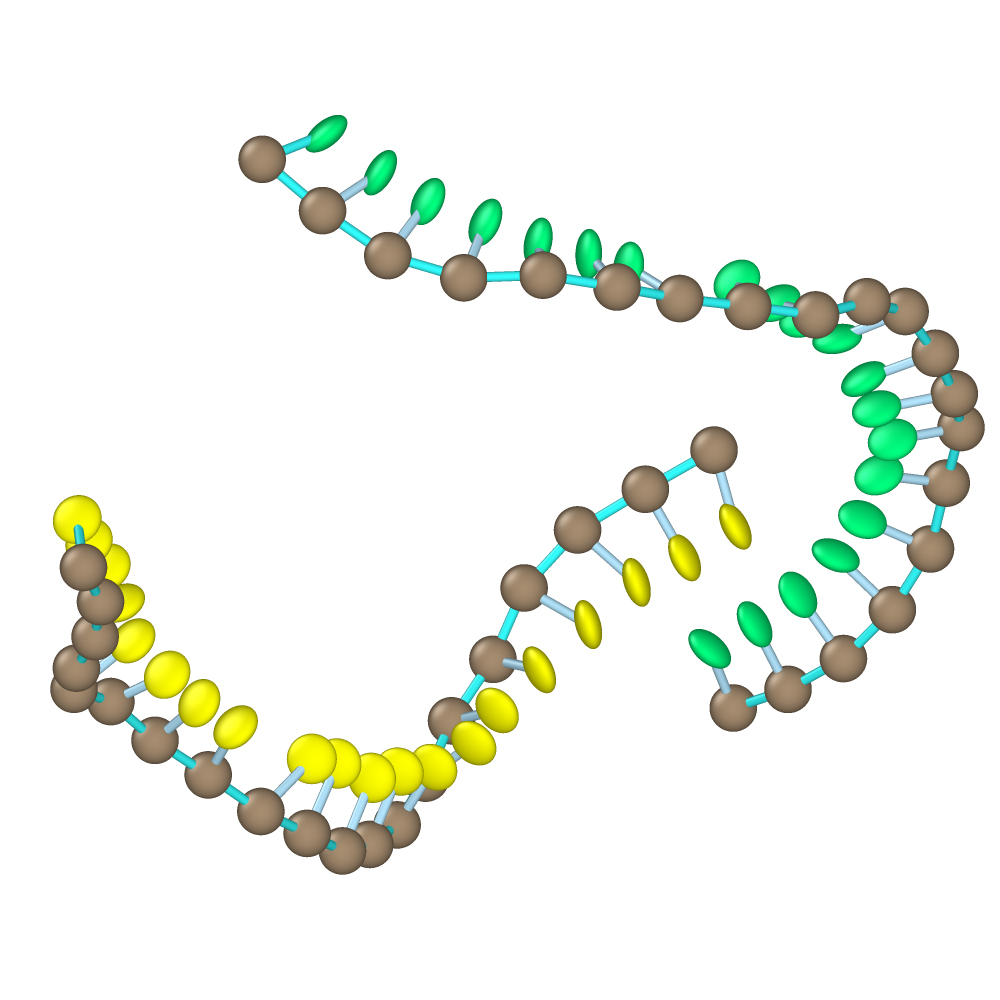}}
        \fbox{ \includegraphics[width=0.4\textwidth]{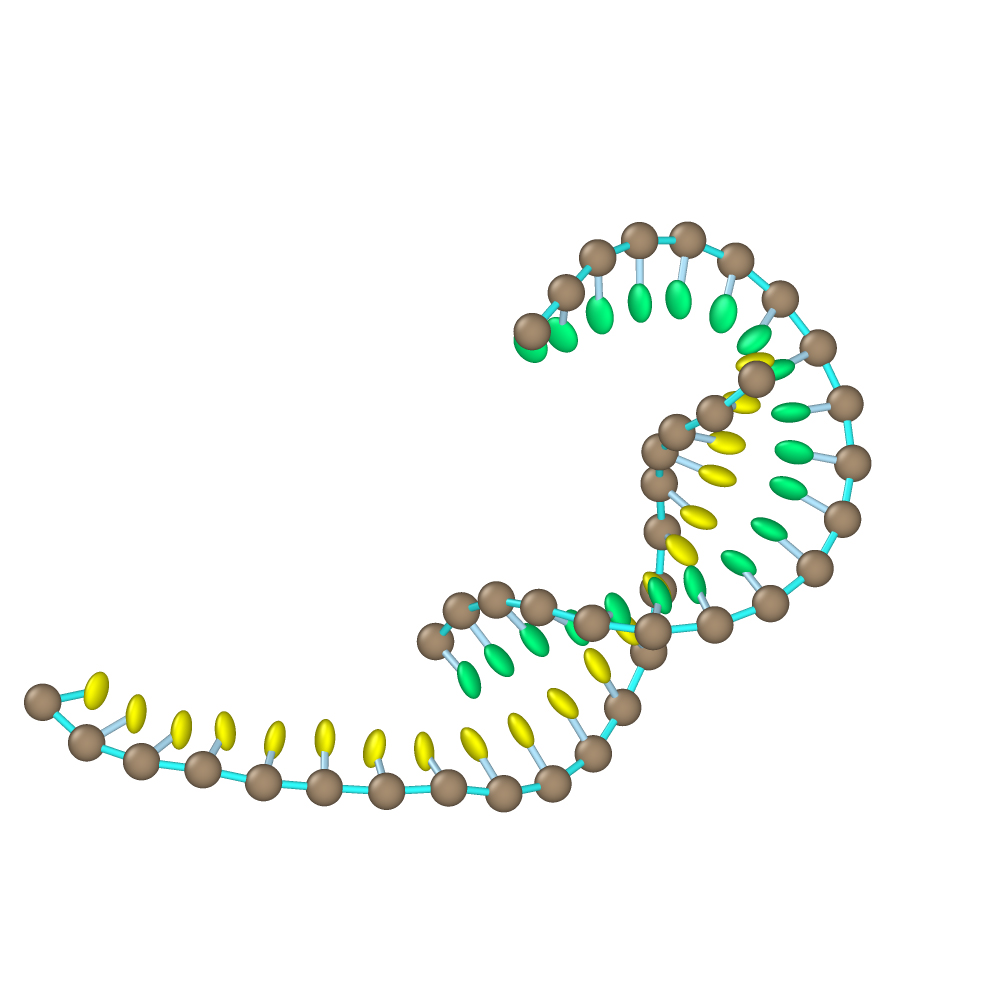}}
        \fbox{ \includegraphics[width=0.4\textwidth]{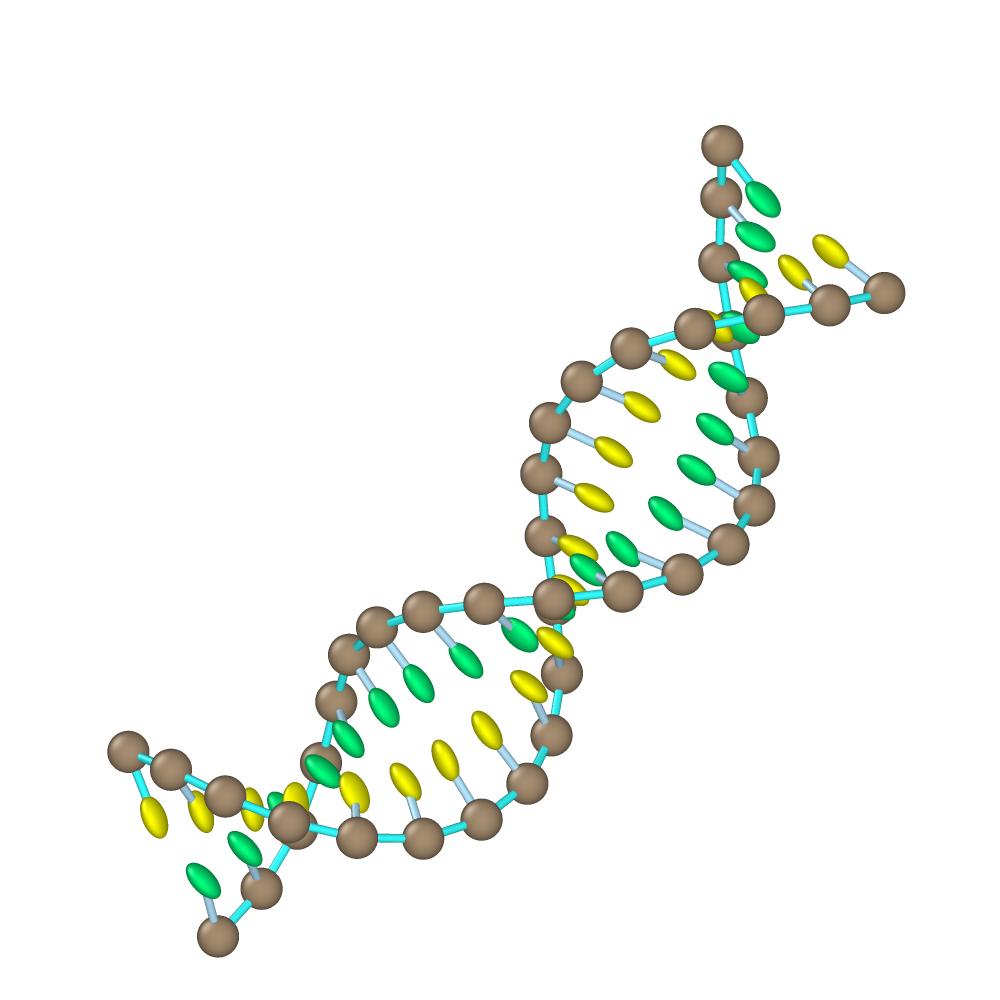}}
        \caption{Snapshots from the simulation. The particle sizes are slightly reduced to enhance visual clarity. (from top-left to bottom-right) (A) Initially the two chains are separate from each other. At around \(T^\ast = 0.8\), the system has lost all memory of its initial configuration and is in a random configuration. (B) As the system is cooled beyond the transition point, the bases begin binding with their complementary counterparts. At around \(T^\ast = 0.4\), the energy barrier has lowered enough for the bases to bind. (C) At around \(T^\ast = 0.35\), The bases in one strand adjust themselves to align with their complementary bases in the other strand. This is feasible since in our model, base pairing lowers effective energy. (D) Finally, at a very low temperature of \(T^\ast = 0.2\), all bases pair successfully and the strands hybridize into a double helix structure.}
        \label{fig:snapshots}
\end{figure*}

Our model is simulated using in house code written in Fortran 90, where the bases are annealed from a random configuration at high temperature to a minimum energy configuration at low temperature. The melting profiles obtained are in qualitative agreement with reference experimental melting profiles, though a full quantitative argument can only be put forward when we improve the parameterization of the potential. In Figure~\ref{fig:snapshots}, we also provide snapshots of the simulation to elucidate the annealing process. However, our generic DNA model still leaves much to be desired. The implicit solvent Morse potential has a significant effect on how the helix forms, but its form and application in the model is incompatible with the formation of unequal groove widths. A more accurate treatment of groove geometry would require either a higher resolution model or a reformulated implicit solvent term. Also, the dihedral potential used here is symmetric, which means \(+\phi\) and \(-\phi\) are equivalent. This is not the case in nature, which has an inherent chirality and prefers the right-handed helix for B-DNA. At present, our model makes no distinction between the two-handed helices, which spontaneously form in the absence of any external forces, even within the same strand when both ends twist opposite to each other, creating kinks. This is an important limitation of our model framework, due to which, our model at present does not yet capture the proper right chirality and detailed helical geometry associated with DNA. Proper incorporation of chirality along with better treatment of implicit solvent effects into our model framework to capture groove formation without introducing steric clashes with existing interactions is something that we aim to address in our future work. We believe that the proper inclusion of such effects will resolve much of the current limitations of our model, as far as the physical properties of DNA are concerned.

\section{Acknowledgements}
A.D. acknowledges the support funded by the Council of Scientific and Industrial Research with grant number (09/0028(19444)/2024-EMR-I). This work is supported by RUSA 2.0 scheme of the University of Calcutta. Computational resources were provided by the University of Calcutta. OpenMP\texttrademark was used for code parallelization. The software OVITO\texttrademark~\cite{ovito} was used for simulation snapshots. The graphs were plotted in the Python programming language using the matplotlib plotting library.

\section{Data Availability}
The data that support the findings of this study are available from the corresponding author upon
reasonable request.

\section{Disclosure Statement}
The authors declare no competing financial interest.

\nocite{das2026coarsegrainedmodelingself}
\printbibliography
\newpage
\appendix
\section*{Appendix A: The soft ECP type potential}

The potential equation for ECP is
similar to that of Gay Berne, given by
\begin{align*}
  U_{GB}(\vec{r}, \hat{\alpha}, \hat{\beta})
  &= 4 \, \epsilon_0\,\epsilon_1^\nu(\hat{\alpha}, \hat{\beta}) \,
  \epsilon_2^\mu(\hat{\alpha}, \hat{\beta}, \hat{r})
  \left[ \rho ^{-12}
    - 2\rho^{-6}
  \right] \\
  \rho &= \frac{r - \sigma(\hat{\alpha}, \hat{\beta},
    \hat{r}) + \sigma_0}{\sigma_0}
\end{align*}
The orthogonal rotation matrices \(\hat{\alpha}\) and \(\hat{\beta}\)
are obtained from the respective rotation quaternion associated with each of the two interacting ellipsoids that we use in
our simulation. 
Since the potential is a pair potential, we represent the two particles using shape matrices \(S_1\) and \(S_2\) given by
\begin{equation*}
S_1=\left(\begin{matrix}\sigma_{1_x}&0&0\\0&\sigma_{1_y}&0\\0&0&\sigma_{1_z}\\\end{matrix}\right);
\qquad
S_2=\left(\begin{matrix}\sigma_{2_x}&0&0\\0&\sigma_{2_y}&0\\0&0&\sigma_{2_z}\\\end{matrix}\right)
\end{equation*}
Here \(\sigma_x, \,\sigma_y\,\) and \(\sigma_z\) denote the respective diameters for the two interacting ellipsoids. This defines the geometry of the particles in the molecular frame. To represent the rotation of the particle, we convert the shape matrix from molecular frame into the lab frame to obtain matrices \(A\) and \(B\) given by
\begin{align*}
      A = \hat{\alpha}^T S_1^2\hat{\alpha} \\
      B = \hat{\beta}^T S_2^2\hat{\beta}
\end{align*}
The matrices \(A\) and \(B\) now encompass the orientational geometry of the ellipsoids. The point of contact must lie on a line of position vectors \( \vec{r} \) satisfying \( A(\vec{r}) = B(\vec{r})\) where \( A(\vec{r}) = (r-r_A)^TA(r-r_A)\) and \( B(\vec{r}) = (r-r_B)^TB(r-r_B)\), \( \vec{r}_A\) and \( \vec{r}_B\) being the respective ellipsoid centers. To find the contact point systematically, Perram et al.\cite{Perram1996} constructed a parametric function of both ellipses \(\varphi\) given by
\begin{equation*}
  \varphi = \lambda_D(1-\lambda_D)\hat{r}[(1-\lambda_D) A + \lambda_D B]^{-1} \hat{r}
\end{equation*}
Where \(r = r_A - r_B\) signifies the direction of approach. This now becomes an optimization problem. The point of contact can be found out from the value of \(\lambda_D\) that maximizes the function \( \varphi \). The shift term is then given by
\begin{equation*}
  \sigma^{-2} = \varphi = \lambda_D(1-\lambda_D)\hat{r}[(1-\lambda_D) A + \lambda_D B]^{-1} \hat{r}
\end{equation*}
All that is left is to construct the function \( \varphi\) itself and find its stationary point using optimization algorithms such as Brent's method~\cite{Brent1973}.
The energy \(\epsilon\) consists of three terms \(\epsilon_0\), \(\epsilon_1\) and \(\epsilon_2\). We set \(\epsilon_0 = \max(\epsilon_x, \epsilon_y, \epsilon_z)\). For \(\epsilon_1(\alpha, \beta)\) we use the formula given by Berardi et al.~\cite{Berardi1995}
\begin{equation*}
  \epsilon_1(\alpha, \beta) = \sqrt{\frac{2s_1s_2}{\textrm{det}(E)}}\,; \qquad s_i = \frac{1}{8} (\sigma_{i_x}\sigma_{i_y} + \sigma_{i_z}^2)(\sigma_{i_x}\sigma_{i_y})^\frac{1}{2} \quad  (\textrm{for }i=1,2)
\end{equation*}
where \(E = A + B\) and \(i=1, 2\) represent the two interacting ellipsoids. This general form is applicable to both identical particles and mixtures.

The term \(\epsilon_2\) depends not only on the orientation but on the inter center distance \(r\). We calculate it similar to the way we calculate \(\sigma \) with slight modifications. The energy matrices \(F_1\) and \(F_2\) in the molecular frame is given by
\begin{align*}
F_1&=\left(\begin{matrix}\left(\frac{\epsilon_0}{{\epsilon_1}_x}\right)^{1/\mu}&0&0\\0&\left(\frac{\epsilon_0}{{\epsilon_1}_y}\right)^{1/\mu}&0\\0&0&\left(\frac{\epsilon_0}{{\epsilon_1}_z}\right)^{1/\mu}\\\end{matrix}\right);
\\
F_2&=\left(\begin{matrix}\left(\frac{\epsilon_0}{{\epsilon_2}_x}\right)^{1/\mu}&0&0\\0&\left(\frac{\epsilon_0}{{\epsilon_2}_y}\right)^{1/\mu}&0\\0&0&\left(\frac{\epsilon_0}{{\epsilon_2}_z}\right)^{1/\mu}\\\end{matrix}\right)
\end{align*}
As before, we convert the diagonal matrices from the molecular frame to the lab frame to obtain matrices M and N which represent the orientation dependent energy strength
\begin{align*}
  M &= \hat{\alpha}^T F_1\hat{\alpha} \\
  N &= \hat{\beta}^T F_2\hat{\beta}
\end{align*}
Similar to the case of \( \varphi\), we construct an optimization problem for the function \(\psi\) with parameter \(\lambda_E\). The stationary point gives us the energy scale of the potential in the respective direction.
\begin{equation*}
  \epsilon_2 = \psi = \lambda_E\,(1-\lambda_E)\,\hat{r}\,[(1-\lambda_E) M + \lambda_E N]^{-1} \, \hat{r}
\end{equation*}
This potential form accurately represents geometry of ellipsoidal molecules, contrary to the Gay Berne implementation. 

The forces and torques for the distance dependent part of a general ellipsoid contact potential is given
by Allen et al.~\cite{Allen2006}. Here we write the expressions for the forces and torques for the soft ECP type potential model. Let \(G = [(1-\lambda_D)A + \lambda_D B]\). Then we
can write
\begin{equation*}
  \varphi = \lambda_D\,(1-\lambda_D)\,\hat{r}\cdot G^{-1} \cdot \hat{r}
\end{equation*}
If we also assume \(\vec{k} = G^{-1} \hat{r}\) such that \(\vec{k}\)
the solution of a linear equation \(G\cdot \vec{k} = \hat{r}\), then
we can write the derivative of \(\varphi \) with respect to
\(\hat{r}\) in terms of \(k\). It is given by
\begin{equation*}
  \frac{\pa \varphi}{\pa \hat{r}} = 2\lambda_D\,(1-\lambda_D)\frac{\vec{k}}{r}
\end{equation*}
The derivative for the term \(\psi\) has the same mathematical
form as that of \(\varphi \) with \(\lambda_D\) replaced by
\(\lambda_E\) and \(\vec{k}\) replaced by \(\vec{k_E}\). With this we
can write the equation of the interparticle force, which is
\begin{align*}
  F &= \epsilon_1^\nu\,\mu\epsilon_2^{\mu-1}\left[ \rho^{-12} -
  \rho^{-6}\right] \frac{2\lambda_E\,(1-\lambda_E)}{r^2} \left[
  \vec{k_E} - (\vec{k_E}\cdot \hat{r}) \hat{r} \right] \nonumber\\ &+
  \epsilon_1^\nu\,\epsilon_2^{\mu} \left[ 12\rho^{-7} - 12
  \rho^{-13}\right] \left( \frac{\vec{r}}{\sigma_0r} +
    \frac{2\lambda_D\,(1-\lambda_D)}{2\sigma_0r^2}\varphi^{-\frac{3}{2}}\,\left[
  \vec{k} - (\vec{k}\cdot \hat{r}) \hat{r} \right] \right)
\end{align*}
We also give the equations for the torques acting on the ellipsoids,
incorporating the derivative of \(\epsilon_1\) with respect to
\(\hat{\alpha}\) and \(\hat{\beta}\) into the respective equation
\begin{align*}
  \tau_1 &= \nu\,\epsilon_1^{\nu-1}\,\epsilon_2^{\mu}\left[
  \rho^{-12} - \rho^{-6}\right] \sum_m \left[ a_m \times E^{-1} \cdot
  a_m \right] \nonumber\\ &+ \epsilon_1^\nu\,\mu\epsilon_2^{\mu-1}\left[
  \rho^{-12} - \rho^{-6}\right]
  \frac{2\lambda_E\,(1-\lambda_E)^2}{r^2} \left[ \vec{k}_E \cdot M
  \times \vec{k}_E \right] \nonumber\\ &+  \epsilon_1^\nu\,\epsilon_2^{\mu}
  \left[ 12\rho^{-7} - 12 \rho^{-13}\right]
  \left(\frac{2\lambda_D\,(1-\lambda_D)^2}{2\sigma_0r^2}\varphi^{-\frac{3}{2}}\,\left[
  \vec{k} \cdot A \times \vec{k} \right] \right)
\end{align*}
and
\begin{align*}
  \tau_2 &= \nu\,\epsilon_1^{\nu-1}\,\epsilon_2^{\mu}\left[
  \rho^{-12} - \rho^{-6}\right] \sum_m \left[ b_m \times E^{-1} \cdot
  b_m \right] \nonumber\\ &+ \epsilon_1^\nu\,\mu\epsilon_2^{\mu-1}\left[
  \rho^{-12} - \rho^{-6}\right]
  \frac{2\lambda_E^2\,(1-\lambda_E)}{r^2} \left[ \vec{k}_E \cdot N
  \times \vec{k}_E \right] \nonumber\\& +  \epsilon_1^\nu\,\epsilon_2^{\mu}
  \left[ 12\rho^{-7} - 12 \rho^{-13}\right]
  \left(\frac{2\lambda_D^2\,(1-\lambda_D)}{2\sigma_0r^2}\varphi^{-\frac{3}{2}}\,\left[
  \vec{k} \cdot B \times \vec{k} \right] \right)
\end{align*}
where \(\epsilon: E^{-1} A = \sum_m \left[ a_m \times E^{-1} \cdot
a_m \right]\) and \(a_m\) are the rows of the matrix \(\hat{\alpha}
S_1\). A similar equation is obtained for the matrix \(B\).

\section*{Appendix B: Effect of the confinement potential}

\begin{figure}[!hb]
  \centering
  \includegraphics[width=0.45\textwidth]{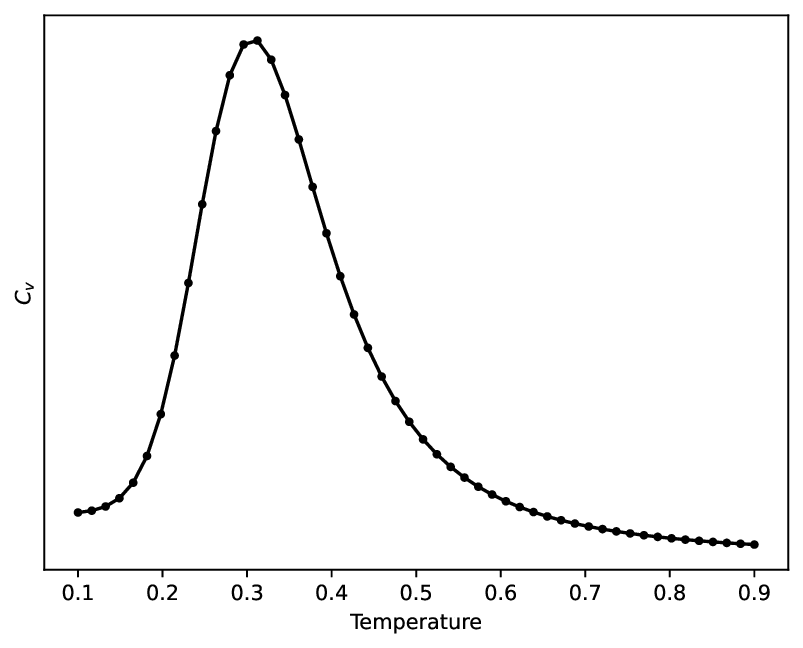}
  \caption{\(C_v\) vs temperature curve plotted using MBAR technique without the harmonic potential turned on. Since we have only two strands in a non-periodic box, presence of the confinement increases the likelihood of hybridization across independent simulation runs.}
  \label{fig:without_well}
\end{figure}

\begin{figure}[!hb]
  \centering
  \includegraphics[width=0.32\textwidth]{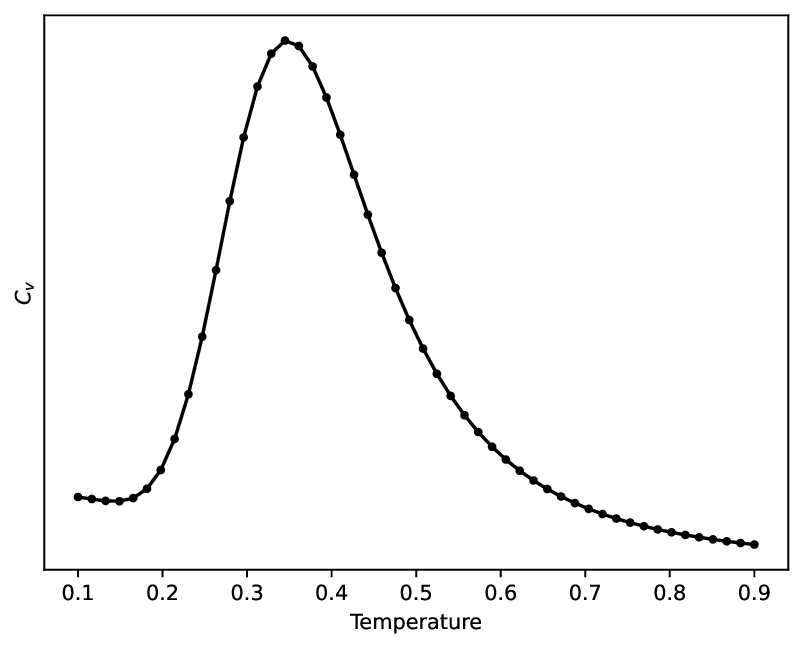}
  \includegraphics[width=0.32\textwidth]{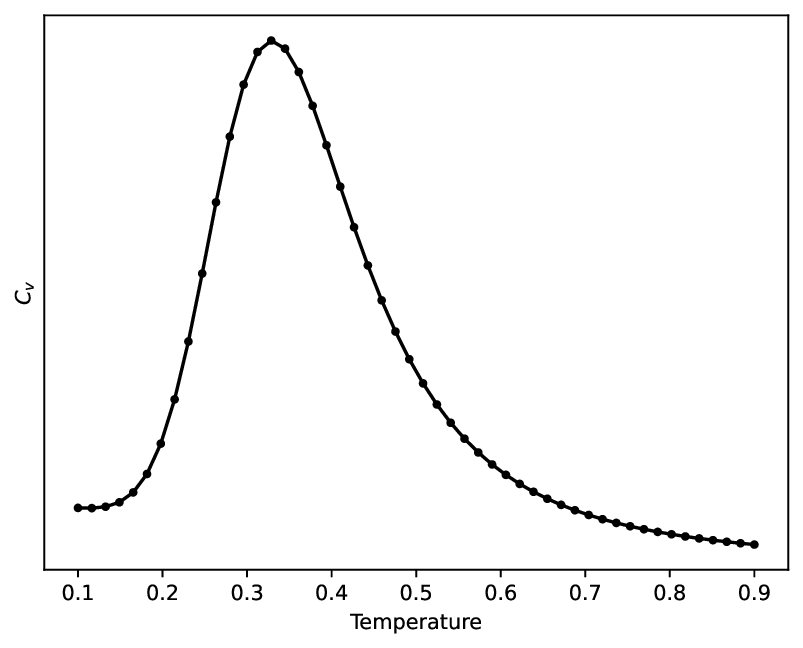}
  \includegraphics[width=0.32\textwidth]{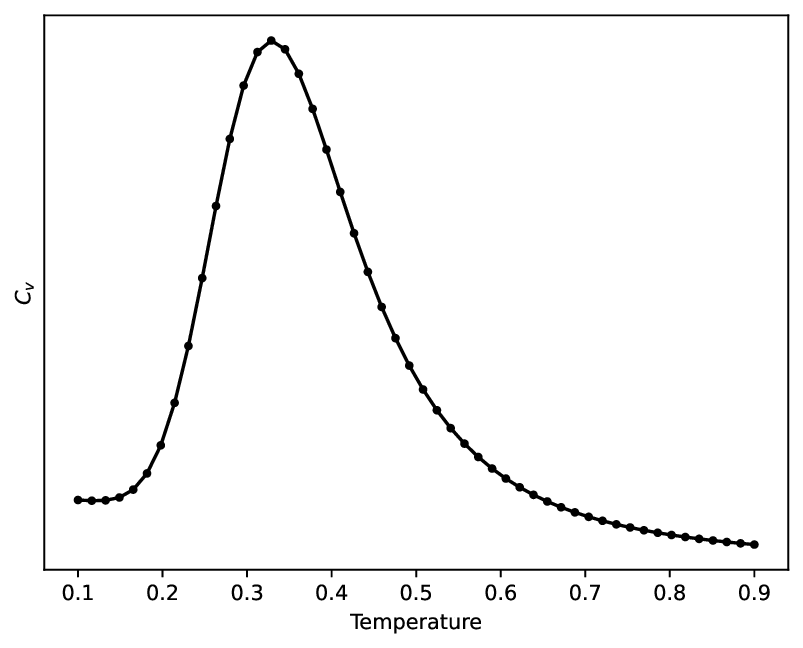}
  \includegraphics[width=0.32\textwidth]{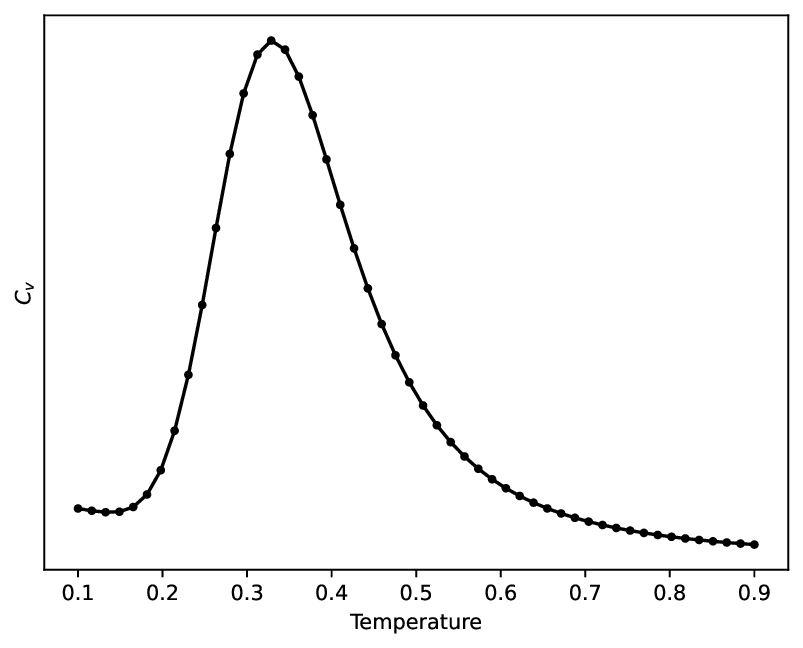}
  \includegraphics[width=0.32\textwidth]{cv_vs_temp_mbar.eps}
  \includegraphics[width=0.32\textwidth]{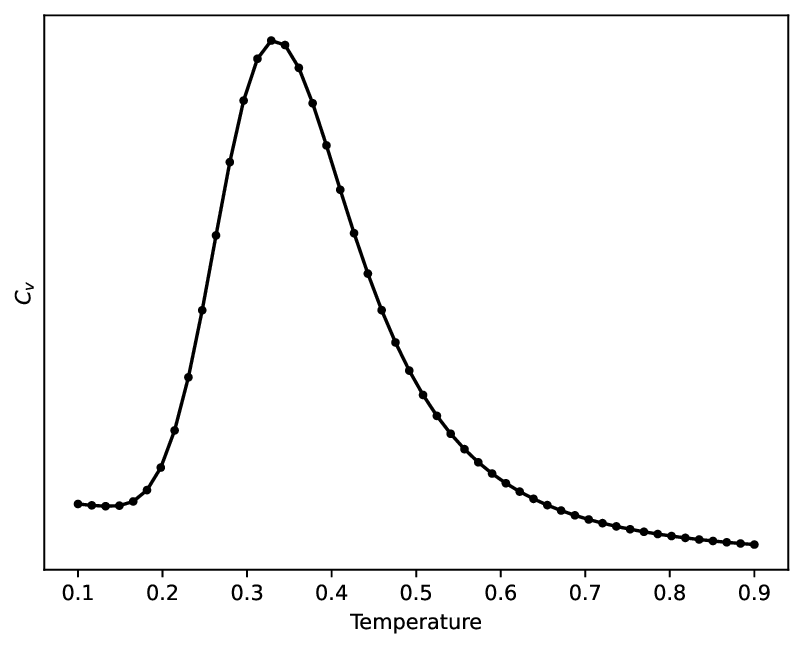}
  \includegraphics[width=0.32\textwidth]{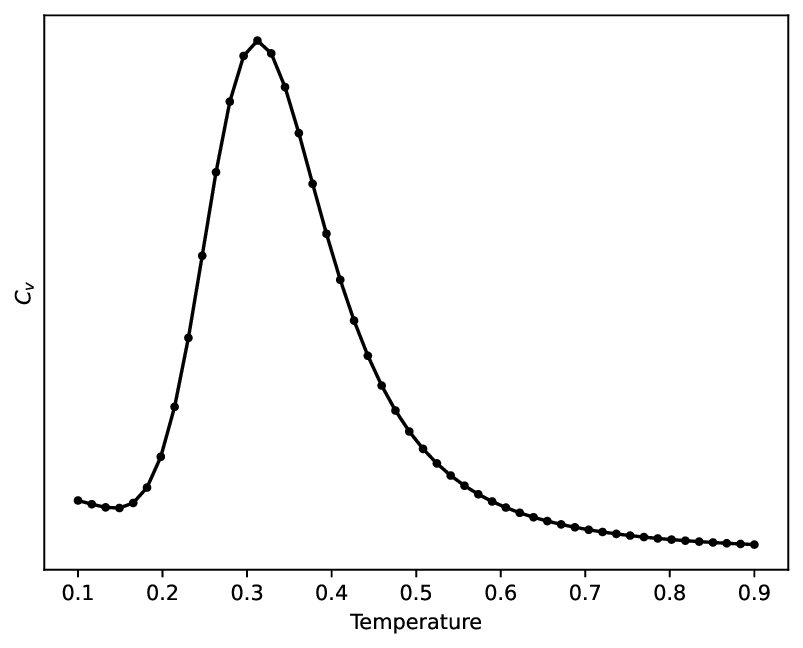}
  \includegraphics[width=0.32\textwidth]{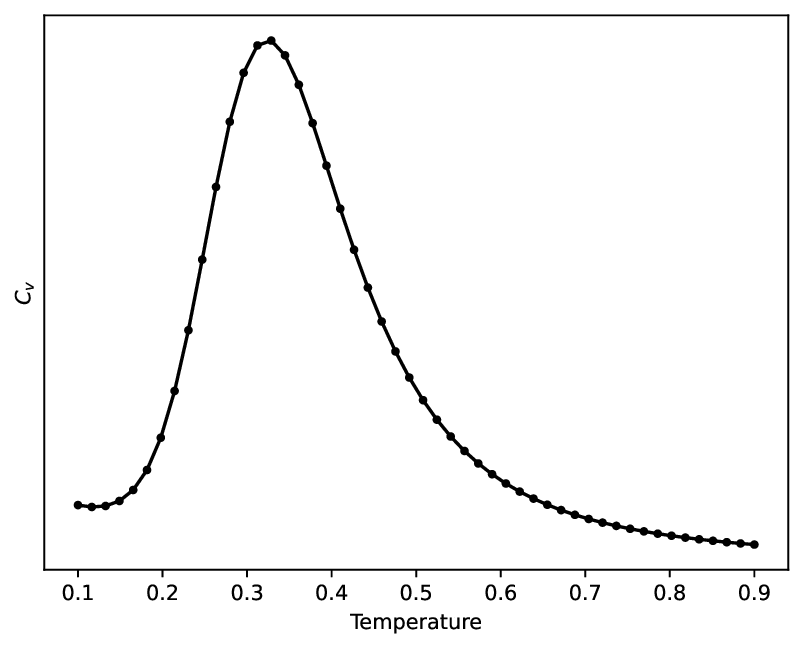}
  \includegraphics[width=0.32\textwidth]{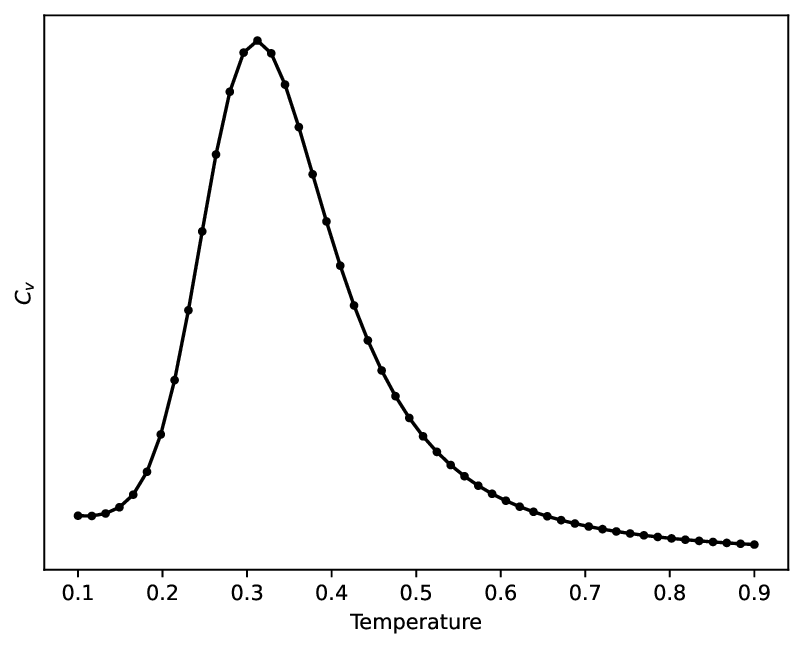}

  \caption{ \(C_v\) vs temperature curve plotted using MBAR technique for different parameters of the confinement potential. There are two parameters for the confinement potential, which are the width of the flat bottom region (\(w_c\)) and the temperature at which it is gradually ramped down to zero (\(\theta_c\)). The first parameter is changed along the vertical direction for values \(w_c\) = \((40, 50, 60)\) from top to bottom in units of \(\sigma_0\). The second parameter is changed along the horizontal direction for values \(\theta_c\) = (0.4, 0.5, 0.6) from left to right in units of reduced temperature. This gives us a set of nine combinations of the two parameters. For example the top left plot has parameters (\(w_c\), \(\theta_c\)) = \((40, 0.4)\) and the bottom right plot has parameters (\(w_c\), \(\theta_c\)) = \((60, 0.6)\). The center plot with parameters \(w_c\) = \(50\) and \(\theta_c = 0.5\) is the one used in the main text. The width of the flat bottom region is chosen to be sufficiently large so that the individual single strands can explore the phase space freely while also ensuring they stay close enough to find each other to form the double helix. Moreover, the restoring force is only applied when the center of the strands diffuse outside the flat bottom radius. This ensures that the two strands effectively sample the configurational phase space while keeping the interaction with the harmonic environment to a minimum. As long as the width of the potential is large enough, and the strands are not entirely at the edge of the region, the system is able to explore the phase space efficiently.}
  \label{fig:confined_cv}
\end{figure}

The flat bottom harmonic potential employed by us introduces a possibility for a bias in the system towards the duplex state. To ensure that we do not introduce improper unphysical bias, we keep the radius of the flat bottom well large and apply the restoring harmonic force only when the center of mass of the individual stands crosses this radius. The restoring force, if applicable, is distributed equally to all atoms in the chain. We do take care to add its contribution to the total potential energy of the system. At around \(T^\ast=0.5\), the harmonic potential strength is ramped down in a sigmoidal fashion until it is almost nonexistent, which is slightly higher than the actual transition point. In our opinion, the effect of the harmonic well would be a loss of configurational entropy of the single stranded state, since the available phase volume is reduced. However, repeated simulations with and without the well (as shown in Fig \ref{fig:without_well}), as well as at different widths for the flat bottom part (as shown in Fig \ref{fig:confined_cv}), show that the effect is negligible. Minor changes in the character of the transition might occur, but since the well effectively ceases to exist close to the transition point, we believe that it is not a major issue.

\end{document}